\documentstyle[aps,prl,epsf,twocolumn]{revtex}                  
\newcommand{\be}[1]{\begin{equation}\label{#1}}
\newcommand{\ee}{\end{equation}}     
\newcommand{\bea}{\begin{eqnarray}}
\newcommand{\eea}{\end{eqnarray}} 
\newcommand{\eq}[1]{Eq.\ (\ref{#1})}
\newcommand{\pic}[2]{\epsfxsize #1 cm
\epsffile{rost-ts#2.eps}} 
\begin{document}  
\tighten   
%
\twocolumn[
\title{ \large\bf Threshold detachment of negative ions by electron 
impact}
\author{  Jan M. Rost}
\address{Wissenschaftskolleg zu Berlin, Wallotstr.~19, D--14193 
Berlin \\ and\\
Fakult\"at f\"ur Physik, Universit\"at Freiburg,
Hermann--Herder--Str.  3,\\ D--79104 Freiburg,
Germany}
\date{\today}
\maketitle
\begin{abstract} \hspace*{1.8cm}\begin{minipage}{142mm}  
The description of threshold fragmentation under long range repulsive 
forces is presented. The  dominant energy dependence near
threshold is isolated by
decomposing the cross section into a product of a back ground 
part and a barrier penetration probability  resulting from the 
repulsive Coulomb interaction.   This tunneling probability contains
the dominant energy variation and it
can be calculated analytically based on the same principles as 
Wannier's description for threshold ionization under attractive 
forces.  Good agreement is found with the available experimental cross 
sections on detachment by electron impact from $D^{-}$, $O^{-}$ and 
$B^{-}$.

\draft
\end{minipage}\end{abstract}

]
Storage ring  based  experiments on threshold detachment from 
the deuteron $(D^{-}$) and the oxygen $(O^{-}$) negative ions by 
electron impact \cite{Andal95,Vejal96,Andal97}, and recently also from $B^{-}$ 
\cite{Andal98}, 
have stimulated  the 
theoretical interest
in the mechanism  and the quantitative description of this process 
\cite{OsTa96,KaTa96,LJL96,Pin96,GrSi98}. 
It is a fundamental question how threshold detachment proceeds since
for very low energies the impacting electron does not even reach the
atom  because it is repelled by the loosely bound electron. 
Early theoretical work on this problem  tried to describe the process
by asymptotic properties of the  wavefunction for the two    electrons
in the continuum after the collision \cite{HGG57}, following the spirit of 
Wigner's treatment for two-body break up \cite{Wig49}.  However, the
predicted cross section agrees purely with the experimental results. 

Some recent theoretical treatments, following another  idea of the early days
\cite{SmCh66}, emphasize   the importance of tunneling contributions,
either  by
treating the impacting electron as a constant perturbing electric 
field \cite{Vejal96}, or by merging a  quantum and a classical 
description \cite{OsTa96}. 
Astonishingly good  agreement with the experiment, 
even  at low energies near threshold,   comes  from
a coupled channel calculation  
in the impact-parameter formalism where
a classical trajectory is used for the relative motion of target and 
projectile electron and the  electron to be detached is described 
quantum mechanically \cite{LJL96}. 
These results, at least the shape of the cross section, depend little 
on the polarization potential used, as Lin etal.\ emphasize  \cite{LJL96}.
 Results of similar accuracy 
have have been reported using  a lowest order distorted-wave 
scheme, however, in contrast to  \cite{LJL96}, with a
 sensitive dependence on the 
polarization potential  \cite{Pin96}.

Without a full calculation of all 
electrons, one cannot avoid to use parameters in one or another way, either directly in 
the simpler models \cite{Andal95,Vejal96}, or indirectly in the more involved calculations 
 modeling  polarization potentials for the loosely bound electron 
 \cite{OsTa96,KaTa96,LJL96,Pin96,GrSi98}. 

The  theoretical work so far remains  inconclusive 
concerning a   dominant mechanism of threshold 
detachment, and the reason  for the seemingly contradicting 
findings concerning the robustness of the results
with respect to changes in the polarization potential is unknown. 

A successful description of near threshold detachment focusing on 
threshold properties should naturally depend very little
on details of the polarization since the long range repulsion between
target and projectile electron dominates. Moreover, such an approach
should uncover a  mechanism for threshold detachment  and
thereby  clarify  the issue of robustness with respect to different
polarization  potentials.

In the following we will show that
  threshold detachment by electrons  can  be described with the
same technique which has lead to the successful (and purely classical) 
description of 
threshold ionization under long range {\it attractive} Coulomb forces, 
pioneered by Wannier \cite{Wan53}.

 However,  in order to learn how to  deal with
repulsive Coulomb forces, one must go back  to a  semiclassical formulation
of threshold ionization  and analyze the reason why Wannier's 
classical treatment  was appropriate. Semiclassically, one may write
the  scattering amplitude in the form \cite{Ros98}
\be{S-matrix}
 f = \sum_{j}\sqrt{{\cal P}_{j}}\exp
 [i\Phi_{j}(E)/\hbar-i\nu_{j}\pi/2],
 \ee
 where the sum runs over all  scattering orbits $j$  which contribute  
 with the weight $\sqrt{{\cal P}_{j}}$. The phase contains the Maslov 
 indices $\nu_{j}$ \cite{Gut90} and the 
 action $\Phi_{j}$ along the orbit which may be expressed as 
 \be{action1}
 \Phi_{j}(E) = \phi_{j}(E){E^{-1/2}}
 \ee
  where  $\phi_{j}(E\to 0) =$ const. \cite{Ros98}. This special form is a consequence of the
homogeneous  Coulomb  interaction. It is crucial for the justification 
of the classical treatment since  $E\to 0$, i.e. approaching 
threshold, and $\hbar\to 0$ have the same effect in \eq{S-matrix}.
If $\Phi_{j}$  is real, which is the case for all classically allowed 
trajectories, one arrives by 
stationary phase approximation 
(for $E\to 0$ or $\hbar \to 0$) at the result 
\be{sigma-class}
\sigma = \sum_{j}{\cal P}_{j} = \sigma_{CL}
\ee 
which sums all individual contributions ${\cal P}_{j}$ of the 
trajectories to the classical cross section $\sigma_{CL}$.

Looking for  the dominant energy dependence of
$\sigma(E\to 0)$ we  decompose the cross section into
\be{prod}
\sigma(E) = \sigma_{B}(E)P(E)
\ee
where $\sigma_{B}(E)$ is a smooth back ground cross section with
$\sigma_{B}(E\to 0) =$const. Wannier showed that the dominant 
energy dependence    $P(E)  = P_{*}$ is contained in a single 
 fixed point orbit $j = *$ \cite{Wan53}. 
Formally, this orbit represents an outgoing 
trajectory with fixed  angle $\theta^{*}= \pi$ between the two electrons and 
 symmetric  distances $r_{1}= r_{2}$ of  electron 1 and 2  from the 
core.  It is convenient to use
hyperspherical coordinates with an overall radius  $r^{2} = 
r_{1}^{2}+r_{2}^{2}$ of the system and the hyperangle 
defined by  $\tan \alpha = r_{1}/r_{2}$.  The symmetric escape orbit
corresponds to increasing hyperradius $r$ and fixed $\alpha^{*} = 
\pi/4$.   The potential energy of the  two electrons interacting
with a core   of charge $Z$ can be written in the form of a Coulomb potential 
with an angular dependent charge, $V = 
C(\alpha,\theta)/r$.   For   $Z > 1/4$  the potential at the fixed 
point  is with  $C_{*}=C(\alpha^{*},\theta^{*}) < 
0$ attractive. Hence the relevant threshold orbit is classically 
allowed with a real action $\Phi_{*}$  and, as sketched above, 
the semiclassical  
scattering  amplitude leads for $E\to 0$ to the classical cross 
section with dominant energy  variation of the form
\be{Pwan}
  P_{CL} (E) = (E/E_{0})^{\beta}
\ee
 as 
derived by Wannier \cite{Wan53}.

On the other hand,  for a  fixed point charge $C_{*}>0$ the 
Coulomb interaction is repulsive. Then, the relevant 
threshold orbit is classically forbidden and represents a tunneling 
trajectory with imaginary action $\Phi_{*}= i\Gamma_*$.   In this case the 
semiclassical cross section does not reduce to the classical one in 
the limit $E\to 0$. Rather, its  major energy dependence results
from  a tunneling mechanism which produces a  Gamow factor
\be{tunnel}
P(E) = \exp[-2\Gamma_*(E)/\hbar].
\ee
Clearly, the threshold cross section is through \eq{tunnel} $\hbar$-dependent. 
Nevertheless,  the important dynamical quantities, namely the tunneling 
action $\Gamma_*$, is still given classically.

In the traditional description of classically allowed threshold 
fragmentation of charged particles the initial configuration is 
unimportant -- the energy dependence of 
\begin{figure}
 \pic{8}{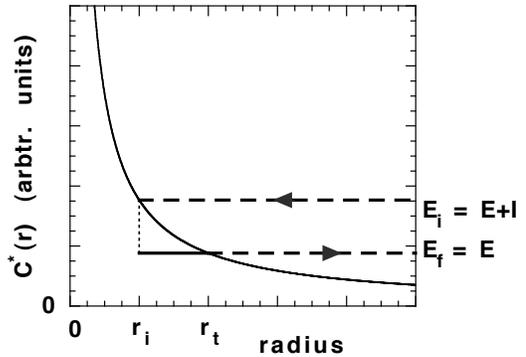}
 \caption[]{Sketch of tunneling threshold dynamics on the fixed point
 manifold with potential $C_{*}(r)$  from \eq{hamil}. The classically 
 allowed  incoming and outgoing trajectory on the respective energies
 $E_{i}$ and $E_{f}$ is shown  (dashed lines) 
 as well as the tunneling part (solid 
 thick line) which determines the threshold fragmentation probability.}
 \end{figure}
 \noindent
the cross section is 
completely determined by the stability of the final fragment 
configuration. This stability enters $P_{*}$ of the escape orbit.
That only the fixed point orbit is relevant close to threshold 
is justified by the fact that  all 
available energy (which approaches zero for $E\to 0$ and $r\to\infty$) must be 
put into the radial degree of freedom $r$ in order to fragment the system.
Hence, the system evolves asymptotically in a frozen configuration 
where
neither its geometrical shape ($\theta = \theta^{*}$), nor the relative
interparticle distances $r_{1}/r_{2} = \tan\alpha^{*}$  change. 
Moreover, due to the Coulomb scaling properties, any   partial wave with 
angular momentum $L$  reduces in scaled coordinates to an $S$-wave
since  the scaled angular momentum reads $\tilde L = L\sqrt{E}$ 
\cite{Ros98}.
Therefore, only the $S$-wave has to be considered which remains
also valid  in the case of a repulsive Coulomb force. Finally,  
for two escaping electrons, the fixed point configuration 
$\theta^{*}= \pi$ and $\alpha^{*} = \pi/4$ remains the same for all 
charges of the core including  the limit  
$Z = 0$  which applies to 
the neutral atom for our problem of electron detachment. Hence, the 
radial motion on the fixed point manifold is governed by the 
Hamiltonian  (atomic units are used unless  otherwise stated)
\be{hamil}
H_{*} = \frac{P_{r}^{2}}{2}+\frac{C(\alpha^{*},\theta^{*})}{r}\,,
\ee
where the effective charge $C_{*} = 2^{-1/2}$  results from 
the evaluation of the electron-electron repulsion $ V = |\vec 
r_{1}-\vec{r_{2}}|^{-1}$ at the fixed point. 

For each energy $E = H_{*}$ 
we can calculate the tunneling action $\Gamma_*(E)$ entering \eq{tunnel}
from the imaginary momentum $p = (-P_{r}^{2})^{1/2}$ of \eq{hamil},
\be{action}
\Gamma_* = \int_{r_{i}}^{r_{t}}p\,dr.
\ee
\begin{figure}
 \pic{9}{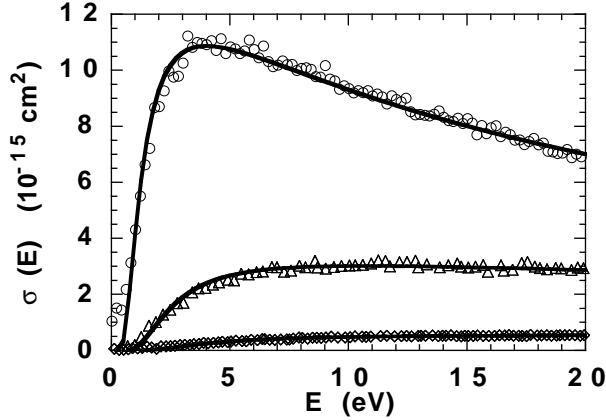}
 \caption[]{Detachment cross section by electron impact as a function of
 excess energy for $B^{-}$ (circles)  from \cite{Andal98},   $O^{-}$ (triangles) and 
 $D^{-}$ (diamonds)  
 from \cite{Andal95}. The  solid lines are the  cross 
 sections from \eq{prod}.}
 \end{figure}
 \noindent
 The integration limits are the outer 
turning  point $r_{t}$ where the orbit becomes classically allowed,
$p(r_{t}) = 0$, 
and a starting point $r_{i}$, see Fig.~1. In  contrast to
threshold fragmentation under attractive Coulomb forces tunneling 
threshold fragmentation depends on the initial configuration, at least 
as far as the value of $r_{i}$ in \eq{action} 
is concerned which will influence shape 
and magnitude of $P(E)$ in \eq{tunnel}.

In a very  crude approximation one could put $r_{i}= 0$ arguing that
the the electronic momentum transfer requires the recoil to be 
absorbed by the nucleus and its position 
is where the outgoing electrons should start.  However, 
in the light of the (small) tunneling probability which determines 
threshold detachment according to 
\eq{tunnel}  close to $E=0$ this is certainly too crude. 
For small excess energy the projectile electron impacts roughly with 
the binding energy $I$ which is of the order of 1 eV. Repelled by the 
loosely bound electron the projectile will never reach $r_{i}\approx 0$
at  this low impact energy. More realistically, one can approximate
$r_{i}$ by the classical turning point of the  incoming electron, 
projected onto the fixed point manifold whose dynamics is specified
by \eq{hamil}. Hence, 
to determine this turning point of the {\it incoming} electron 
we put  $P_{r} = 0$  in \eq{hamil} 
at the incoming electron  energy of $E_{i} = 
E+I$ to yield
\be{ri}
 r_{i} = C_{*}/(E+I).
 \ee
 The initial momentum of the {\it outgoing} electron pair $p(r_{i}) = 
 \sqrt{2I}$ follows from the Hamiltonian  \eq{hamil}  on the final
 energy surface $E_{f}= E$. The situation is sketched in 
 Fig.~1.
 Using \eq{action} and \eq{ri} 
 the threshold detachment probability \eq{tunnel} reads in 
 dimensionless units explicitly
 \begin{figure}
 \pic{9}{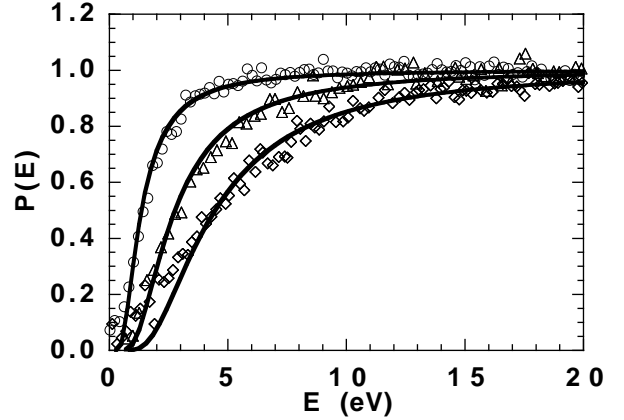}
 \caption[]{Experimental detachment probabilities, obtained
 by dividing the cross section by $\sigma_{B}(E)$ from \eq{backg}. 
 The coding of the data is as in Fig.~3. Theoretical $P(E)$ from 
 \eq{tunnel}.}
 \end{figure}
 \noindent
 \be{prob}
 P(E) = 
 \exp
\left[-4C\alpha\sqrt{\frac{m_{e}c^{2}}{
2E}}\left(\arctan\sqrt{\frac{I}{E}}-\frac{\sqrt{IE}}{I+E}\right)\right],
 \ee
where $\alpha = 1/137$  is the fine structure  constant, $m_{e}c^{2} = 
511$keV is the rest mass of the electron, and  $C = C_{*}$ is the 
repelling charge of the  two electrons on the fixed point manifold in units of
$e$,
see
\eq{hamil}.  One can cast \eq{prob} into a more familiar form
of atomic units by noting that $m_{e}c^{2}/\alpha^{2} = e^{2}/a_{0} =
27.2116$eV is  just the atomic energy unit. Clearly,   the tunneling mechanism
breaks the scaling  invariance of $P(E)$ for different systems characterized 
by different ionization potentials $I$ since  $P(E)$ does not only 
depend on $E/I$ but also on $m_{0}c^{2}/E$. This is one of the  major
differences compared to Wannier's classical result (\eq{Pwan}) for threshold 
ionization under attractive Coulomb forces. 

Different $P(E)$ are shown 
 in Fig.~3 with solid lines  corresponding to detachment from the ions $B^{-},
D^{-}$, 
 and $O^{-}$ respectively.   
 The   ``experimental'' tunneling probabilities are extracted by 
 fitting the experimental cross sections (Fig.\ 2)  to \eq{prod} 
 with
\be{backg}
\sigma_{B}(E) = \sigma_{0}/(b_{0} + E/I),
\ee
 where
 $\sigma_{0}, b_{0}$ are fitting parameters. The $\sigma_{B}(E)$ 
 obtained in this way are shown  in Fig.\ 4 for completeness and 
 exhibit the expected monotonically decreasing behavior. 
 
 As a final support for   the analytical $P(E)$ from \eq{prob} 
 we have  fitted the experimental 
 cross sections with $\sigma_{0},b_{0}$ and $I$ as free parameters. 
 The result for $I$ was  $0.297\pm 0.008, 0.79\pm 0.03$ and 
 $1.58\pm 0.04$eV which is close to the
 accurate values of $0.28, 0.75$ and $1.46$eV for $B^{-},D^{-}$, and 
 $O^{-}$, respectively.  
 \begin{figure}
 \pic{9}{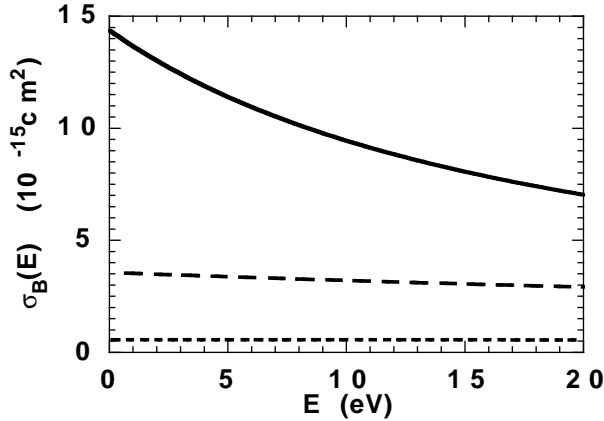}
 \caption[]{Back ground cross sections $\sigma_{B}(E)$ for $B^{-}$ 
 (solid), $D^{-}$ (dashed) and  $O^{-}$ (dotted), see text.}
 \end{figure}
\noindent

 The present description differs from various published tunneling models
 approximating in one or another way the actual electron motion 
  by tunneling. 
 In the present treatment, only the dominant energy dependence of
 the cross section is derived from a fixed point   orbit which 
 represents a tunneling trajectory. However, this trajectory does not 
 correspond to a  true, physical two electron orbit. Rather, 
 it is  a stationary point
 solution for $\hbar\to 0$,  in complete analogy to 
 Wannier's solution for the classically allowed case of attractive 
 forces.  This stationary point  calculated in the limit $E\to 0$ 
 does not depend at all 
 on the polarization potential.
   Only the 
  binding energy of the target 
 electron 
 enters $P(E)$ through $r_{i}$ as defined in \eq{ri}  from the turning 
 point of the incoming trajectory. 
 It is exactly this element which is 
 similarly 
 contained in the  calculation of Ref.\ \cite{LJL96}. 
Hence,  this impact parameter calculation captures an essential feature
of the threshold detachment 
 dynamics  making the whole calculation robust against 
 details of the   polarization potential. These details will 
 influence on the other hand  the background cross section 
 $\sigma_{B}(E)$ much more strongly.  The distorted wave calculation 
 \cite{Pin96}
   by nature approximates the threshold region 
   from an expansion of the high energy limit which is much more 
   sensitive on details of the (shorter range) polarization potential.

 In summary, separating  the rapidly changing 
 detachment probability $P(E)$ from the background cross section 
 $\sigma_{B}(E)$ we have shown that threshold fragmentation under
 asymptotic repulsive Coulomb forces can be treated on the same 
 footing as the well established threshold ionization under 
 attractive Coulomb forces. In contrast to
 the classical result for 
 attractive forces, threshold detachment of negative ions by 
 electrons can be interpreted 
to proceed
  via quantum mechanical tunneling 
of the outgoing electron pair. This implies a breaking 
of the scale 
invariance of $P(E)$ with respect
 to energy  since $P(E)$ depends on $m_{e}c^{2}/E$ irrespectively of the 
 target properties while  $P_{CL}(E)$ (\eq{Pwan}) under attractive Coulomb 
 forces is scale invariant. 
 Yet,  $P(E)$ for threshold detachment can be  described semiclassically
 due to the dominant (repulsive)  Coulomb interaction which ensures
 through its scaling properties that $E\to 0$ also means $\hbar \to 
 0$ (see \eq{action1}).
  The same scaling properties also reduce the 
 dominant energy dependence of all partial waves to that of $L = 0$.
 Therefore,   $P(E)$ can be determined from the S-wave only, as it
 has been done in the present work.

I would like to thank  L.~H.~Andersen for making available the 
experimental data electronically, in the case of $B^{-}$ even prior 
to publication. Financial support from the  DFG under the 
 Gerhard Hess-Programm is  gratefully acknowledged.



 %
\end{document}